# Water Production Rate of C/2020 F3 (NEOWISE) from SOHO/SWAN over Its Active Apparition

M.R. Combi[1], T. Mäkinen[2], J.-L. Bertaux[3], E. Quémerais[3], S. Ferron[4]

Contact email: mcombi@umich.edu

[1]Dept. of Climate and Space Sciences and Engineering
University of Michigan
2455 Hayward Street
Ann Arbor, MI 48109-2143
*Corresponding author: mcombi@umich.edu

[2]Finnish Meteorological Institute, Box 503
SF-00101 Helsinki, FINLAND

LATMOS/IPSL
Université de Versailles Saint-Quentin
11, Boulevard d'Alembert, 78280, Guyancourt, FRANCE

[4]ACRI-st, Sophia-Antipolis, FRANCE

Short title: **Water Production Rate of C/2020 F3 (NEOWISE)**

**ABSTRACT**

C/2020 F3 (NEOWISE) was discovered in images from the Near Earth Object program of the Wide-Field Infrared Survey Explorer (NEOWISE) taken on 27 March 2020 and has become the Great Comet of 2020. The Solar Wind ANisotropies (SWAN) camera on the Solar and Heliospheric Observatory (SOHO) spacecraft, located in a halo orbit around the Earth-Sun L1 Lagrange point, makes daily full-sky images of hydrogen Lyman-alpha. Water production rates were determined from the SWAN hydrogen Lyman-alpha brightness and spatial distribution of the comet measured over a 4-month period of time on either side of the comet's perihelion on 3 July 2020. The water production rate in $s^{-1}$ was moderately asymmetric around perihelion and varied with the heliocentric distance, r, in au as $(6.9\pm0.5) \times 10^{28} r^{-2.5\pm0.2}$ and $(10.1\pm0.5) \times 10^{28} r^{-3.5\pm0.1}$ before and after perihelion, respectively. This is consistent with the comet having been through the planetary region of the solar system on one or more previous apparitions. A water production rates as large as $5.27 \times 10^{30}$ $s^{-1}$ were determined shortly after perihelion, once the comet was outside the solar avoidance area of SWAN, when the comet was 0.324 au from the Sun.

1. **Introduction**

Comet C/2020 F3 (NEOWISE), hereafter comet NEOWISE, is an Old Long-Period comet, using the A'Hearn et al. (1995) classification, discovered in images taken on 27 March 2020 by the Near Earth Object program of the Wide-Field Infrared Survey Explorer (NEOWISE). Its orbit had an inbound semi-major axis of 270 au and is projected to have an outbound semi-major axis of 255 au. It reached perihelion on 3 July at a distance of 0.295 au from the Sun. It became a spectacular visual naked-eye object for a few weeks near perihelion. Its semi-major axis indicates that it is definitely not a dynamically new comet directly from the Oort cloud and so has been through the planet region of the solar system on previous apparitions.

The Solar Wind ANisotropies (SWAN) camera on the Solar and Heliospheric Observatory (SOHO) spacecraft observed the Lyman-alpha emission from atomic hydrogen throughout its apparition. SWAN makes all-sky observations of the Lyman-alpha emission of interstellar atomic hydrogen that streams through the solar system and is removed by solar radiation and solar charged particle impact leaving a characteristic distribution that is bright in the direction of incoming hydrogen atoms and fainter in the downstream direction of outgoing hydrogen atoms (Bertaux et al. 1995). The typical interplanetary hydrogen

Lyman-alpha brightness ranges from about 0.5 kilorayleighs up to several kilorayleighs in directions near the Sun.

SWAN's 5 by 5 one-degree detectors scan across the sky daily, making a full sky image, excluding an avoidance region around the Sun and another region blocked by parts of the spacecraft itself. SOHO has been located in a halo orbit around the L1 Earth-Sun Lagrange point since a few months after its launch in December of 1995. Most instruments are still operating nominally, including SWAN. A comet is detectable when its hydrogen Lyman-alpha brightness is larger than about 100 Rayleighs so it is distinguishable by its spatial distribution above that of the interplanetary signal. This typically happens for comets that are brighter than between magnitude 12 to 10 in the visible. Cometary signals as bright as 20-30 kilorayleighs, which do not saturate the detectors, can be reliably detected. Observations at the faint end are typically limited by the interplanetary brightness, especially near the Sun, and by the occurrence of nearby bright stars, especially when a comet is near the galactic equator.

## 2. Model Analysis

Observations of a comet's hydrogen coma are analyzed in order to calculate the water production rate of the comet, relying on the fact that water is typically the most abundant volatile gas in the coma when comets are within about 3 AU from the Sun (Mäkinen and Combi 2005; Combi et al. 2005). Water is photodissociated into a well-known distribution of possible fragment H and O atoms, $H_2$ molecules, OH radicals and their ions (Combi et al. 2004). Since SOHO's launch in 1995 water production rates have been determined from SWAN observations of over 70 comets (Bertaux et al. 1998; Combi et al. 2019, 2020a).

Comet NEOWISE was detectable and first distinguishable from background stars in SWAN images beginning on 12 May 2020. It was detected from then until 19 June when it entered the solar avoidance area. Enough of the coma to be useful for analysis was clearly detectable beginning on 6 July and remained detectable until 2 September.

Figure 1 shows an image of the H Lyα coma observed by SWAN on 9 July 2020 when it was near its maximum. Since the solar fluorescence rate of hydrogen Lyman-α is rather large, radiation pressure on the hydrogen atoms produces what is essentially a broad hydrogen "tail" pointed away and lagging behind the direction of the comet in its orbit around the Sun. The extent of the asymmetry of and the general shape of the hydrogen coma is determined by velocity distributions of H atoms leaving the inner partially collisional coma (Combi & Smyth 1988; Combi et al. 2000). While H atoms are produced at a dominant velocity of ~18 km s$^{-1}$ upon photodissociation of water and ~8 km s$^{-1}$ upon photodissociation of OH, when coma production rates are large, and especially for a smaller heliocentric distance when photochemical lifetimes are smaller, collisions of the fast nascent hydrogen atoms with the slow outflowing heavy molecules partially thermalizes and slows part of the distribution of hydrogen atoms. For a large production rate comet like C/2020 F3 (NEOWISE) at small heliocentric distances (0.3-0.5 au) a significant fraction of the hydrogen atoms are slowed to velocities from 8-18 km s$^{-1}$ to 1-4 km s$^{-1}$, elongating and narrowing the antisunward distribution of the hydrogen tail. Figure 2 shows the profile cut through the hydrogen coma indicated by the straight line in Figure 1. In the model that we use, the nascent velocity distribution is determined solely by the known photodissociation branches of water and OH, and the velocity distribution of the atoms exiting the inner partially collisional coma is determined by the level of the water production rate. The velocity distribution then is not a fitting parameter, but a consequence of the level of the production rate and the heliocentric distance. Therefore, the fact that the model reproduces the spatial distribution of the coma is a demonstration that the fixed parameterization of the coma physics is correct given the linear relationship between the observed brightness and abundance as well as the non-linear relationship between the spatial distribution of the brightness and the photochemical and collisional physics in the model.

Water production rates were calculated from all hydrogen coma images using the method described in detail by Mäkinen and Combi (2005). Water production rates were calculated from each SWAN image of comet F3 (NEOWISE) for 74 days beginning on 13 May 2020 and running through 2 September 2020. There was a large data gap from 18 June until 7 July when the comet was too close to the Sun in the sky for SWAN to obtain images of the region of the sky where the comet was located. The first post-perihelion production rate obtained when the comet was on 0.324 au from the Sun on 7 July was 5.15 x 10$^{30}$ s$^{-1}$, one of the largest water production rates obtained by SWAN since comet C/1995 O1 (Hale-Bopp) (Combi et al. 2000). For Hale-Bopp Combi et al. (2000) explained that for a brightness less than ~30 kilorayleighs optical depth effects in the SWAN aperture could be neglected. For F3 (NEOWISE) the brightest pixel centered on the nucleus has a comparable brightness, but this is only because the smallest heliocentric distance is ~0.33 au when the comet was nearly this bright and the illuminating flux from the Sun ~8 times larger than it was for Hale-Bopp at perihelion and column densities similarly many times lower. Therefore, optical depth effects for F3 (NEOWISE) are much less of an issue.

Table 1 gives the observational circumstances, g-factors, the water production rates, and their formal 1-sigma uncertainties. Figure 3 shows the variation of the water production rate as a function of time in days measured from perihelion on 3 July 2020. The vertical lines through each point give the formal 1-sigma error bars that result from the fitting procedure to the hydrogen coma model and the interplanetary hydrogen Lyman-alpha sky background. Actual systematic uncertainties resulting from the model, model parameters and the absolute calibration of SWAN, the LASP solar Lyman-alpha irradiances and the solar Lyman-alpha line profile are estimated to be on the order of 30%. It appears that the water production rate at perihelion when SWAN could not observe it was on the order of nearly $10^{31}$ s$^{-1}$.

## 3. Discussion

It is often instructive to look at a power-law fit of the water production rate as a function of heliocentric distance both before and after perihelion. Figure 4 shows the water production rate plotted as a function of heliocentric distance with pre- and post-perihelion data plotted in separate panels. The straight lines are the best-fit power laws. In the survey of 61 comets observed by SOHO/SWAN, Combi et al. (2019) have shown there is a consistent pattern of variation of the average and range of pre- and post-perihelion power-law exponents as a function of the dynamical age of long period comets, with the flattest slopes found in Dynamically New comets ($1/a_o < 50 \times 10^{-6}$) and gradually more steeper slopes in Young Long-Period ($50 \times 10^{-6}$ $1/a_o < 2000 \times 10^{-6}$) and Old Long-Period comets ($1/a_o > 2000 \times 10^{-6}$), using the A'Hearn et al. (1995) dynamical age classification. The power-law fits to the water production rates are $(6.9\pm0.5) \times 10^{28}$ $r^{-2.5\pm0.2}$ and $(10.1\pm0.5) \times 10^{28}$ $r^{-3.5\pm0.1}$ before and after perihelion, respectively, with the production rate in s$^{-1}$ and the heliocentric distance, r, in au. The pre-perihelion power-law exponent determined for C/2020 F3 (NEOWISE) is near the middle of the range of comets, in the same dynamical Old Long-Period classification from the survey. The post-perihelion exponent is also consistent, being in the same range with comets that are similarly Old Long-Period in the survey. The deviations of the actual production rate determination from the power-law fits are surprisingly small, showing that there were no major outbursts or quiescent episodes.

Bauer et al. (2020) have estimated a nucleus diameter of ~5 km for this comet from WISE spacecraft observations by subtracting a fitted dust coma model. This means that the nucleus is only slightly larger than that of the Rosetta target comet 67P/Churyumov-Gerasimenko, which had a mean diameter of ~4 km. The water production rate of 67P was in the range of 1–3 x 10$^{28}$ s$^{-1}$ (Bertaux et al. 2014, Biver et al. 2019; Combi et al. 2020b) when it reached its perihelion of 1.24 AU. When C/2020 F3 (NEOWISE) was at a heliocentric distance of 1.2 — 1.3

AU its water production rate was in the range of $4 - 6 \times 10^{28}$ s$^{-1}$ (Table 1). Whereas the production rate of NEOWISE was a factor 2-4 times larger than that of 67P at the same heliocentric distance, it's surface area is about 56% larger. So, while 67P had an active surface fraction in the range of 4 — 8%, that of NEOWISE was on the order of 8 — 20%.

## 4. Summary

Comet C/2020 F3 (NEOWISE) was a spectacular naked object in late June and early July 2020. The SOHO/SWAN all-sky camera observed its Lyman-alpha hydrogen coma on 74 days during the period from 13 May through 6 September 2020. When it was recovered after exiting the SOHO solar avoidance area a few days after perihelion, the hydrogen coma displayed a broad and extended tail produced by solar Lyman-alpha radiation pressure that was over 15° long. This was well reproduced by our analysis model that accounts for the nascent velocity distribution and the partial collisional thermalization (and slowing) of atoms as they exit the inner coma. This model was used to calculate water production rates for each image. A maximum water production rate of $5.27 \times 10^{30}$ s$^{-1}$ was recorded on 8 July when the comet was at a heliocentric distance of 0.337 au. While this was one of the largest water production rates ever recorded by SOHO/SWAN (Combi et al. 2019), once the comet was at a heliocentric distance of 1.2-1.3 au, comparable to the perihelion measurements of the Rosetta target comet 67P/Churyumov-Gerasimenko, its production rate was only a factor of ~3 larger. This, combined with its determined radius of ~5 km, compared with that of 67P of ~4 km, implied that its active fraction was only about a factor of 2 larger than 67P, a rather typical Jupiter family comet.

The orbit of F3 NEOWISE, having an inbound semi-major axis of 270 au, implies that it is an Old Long-Period comet, indicates that it has been through the planetary region of the solar system at least once, and possibly many times. The heliocentric distance dependence, described by power-law fits of production rates as a function of heliocentric distance of $(6.9\pm0.5) \times 10^{28}$ $r^{-2.5\pm0.2}$ and $(10.1\pm0.5) \times 10^{28}$ $r^{-3.5\pm0.1}$, respectively, before and after perihelion, are consistent with this dynamical age. There were no noticeable outbursts, quiescent periods, or evidence of any fragmentation throughout the time covered by the SWAN observations.

**ACKNOWLEDGEMENTS:** SOHO is an international mission between ESA and NASA. M. Combi acknowledges support from NASA grant 80NSSC18K1005 from the Solar System Observations Program. T.T. Mäkinen was supported by the Finnish Meteorological

Institute (FMI). J.-L. Bertaux and E. Quémerais acknowledge support from CNRS and CNES. We obtained orbital elements from the JPL Horizons web site (http://ssd.jpl.nasa.gov/horizons.cgi) and the Minor Planets Center https://www.minorplanetcenter.net. The composite solar Lyα data was were taken from the LASP web site at the University of Colorado (http://lasp.colorado.edu/lisird/lya/). We acknowledge the personnel that have been keeping SOHO and SWAN operational for over 20 years, in particular Dr. Walter Schmidt at FMI.

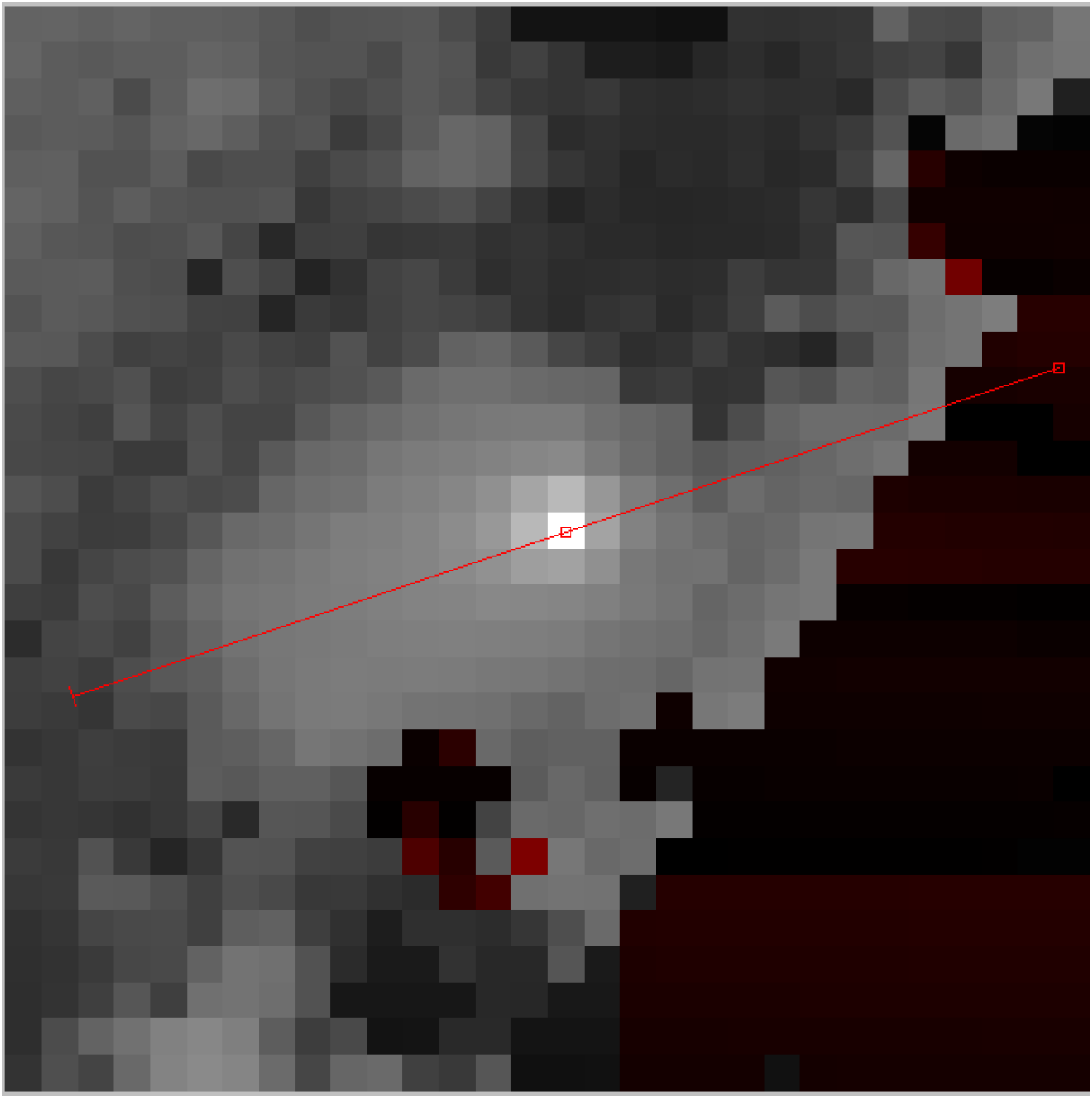

Figure 1. SOHO/SWAN image of the hydrogen coma of comet C/2020 F3 (NEOWISE) observed on 9 July 2020. The field of view is 30° across. The straight line from left to right and slightly upward shows a profile cut along the maximum extent of the hydrogen "tail" whose profile is shown in Figure 2. The dark region toward the right and lower right corresponds to the solar avoidance area of SWAN with the direction to the to the right Sun.

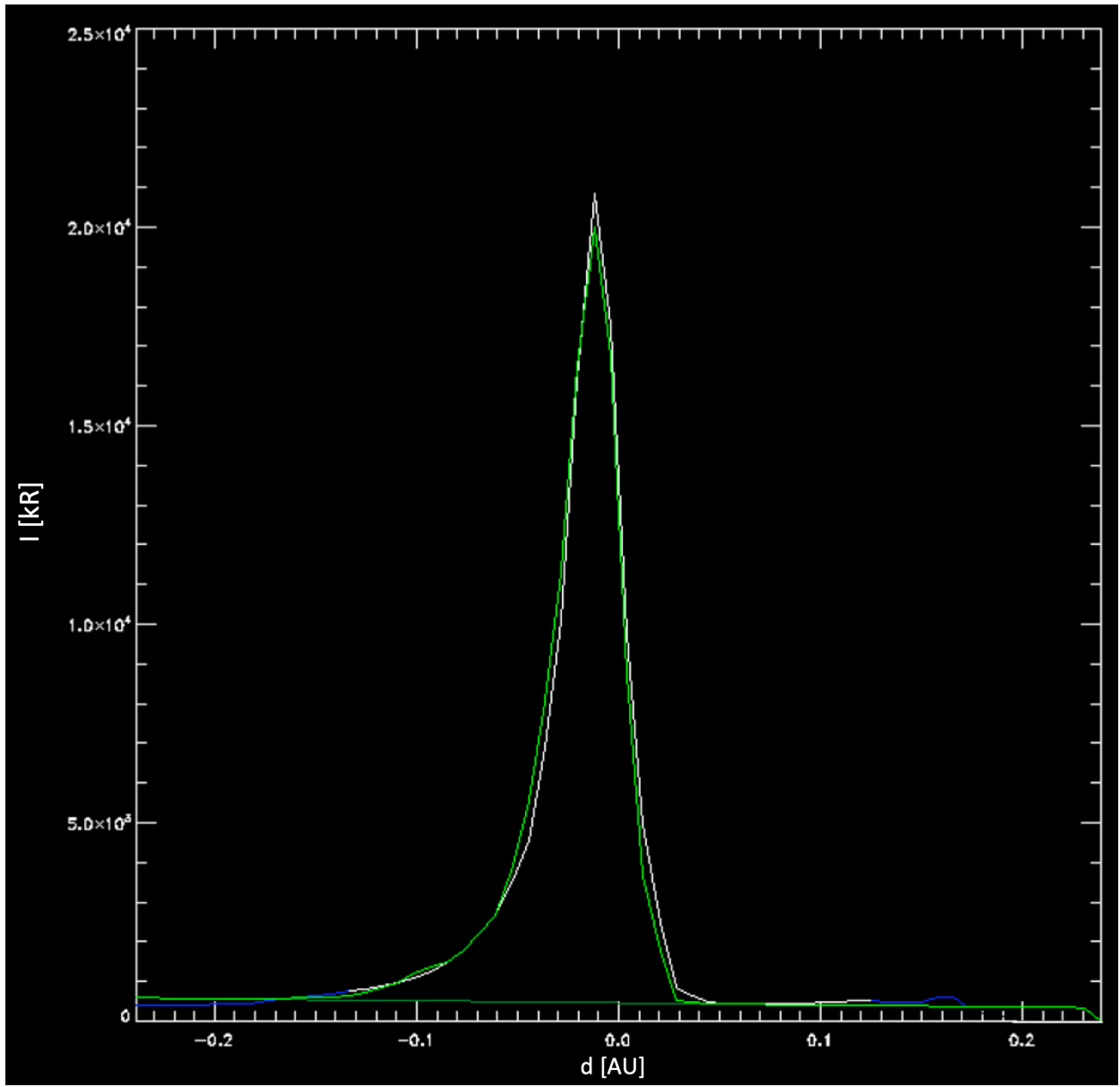

Figure 2. Model-Data comparison profile along the long extent of the hydrogen "tail" of comet C/2020 F3 (NEOWISE) observed by SOHO/SWAN on 9 July 2020. The white curved line corresponds to the observed brightness along the cut shown in Figure 1 along the maximum extent of the hydrogen tail. The flat straight green line near the bottom corresponds to the fitted interstellar hydrogen background. The curved green line is the fit of the coma model to the observed image. The red line toward the lower right is where the observed profile cuts through a field star that has been marked to be ignored. Note that the shape of the model profile follows the asymmetric coma distribution that is produced by radiation pressure from the resonance scattering of the same solar Lyman-alpha photons that enable us to observe the hydrogen coma. The asymmetric shape itself is a property of the model parameterization itself and not a free fitting parameter. The fact that it reproduces the data indicates that the model works well even under these extreme conditions of high production rate and small heliocentric distance.

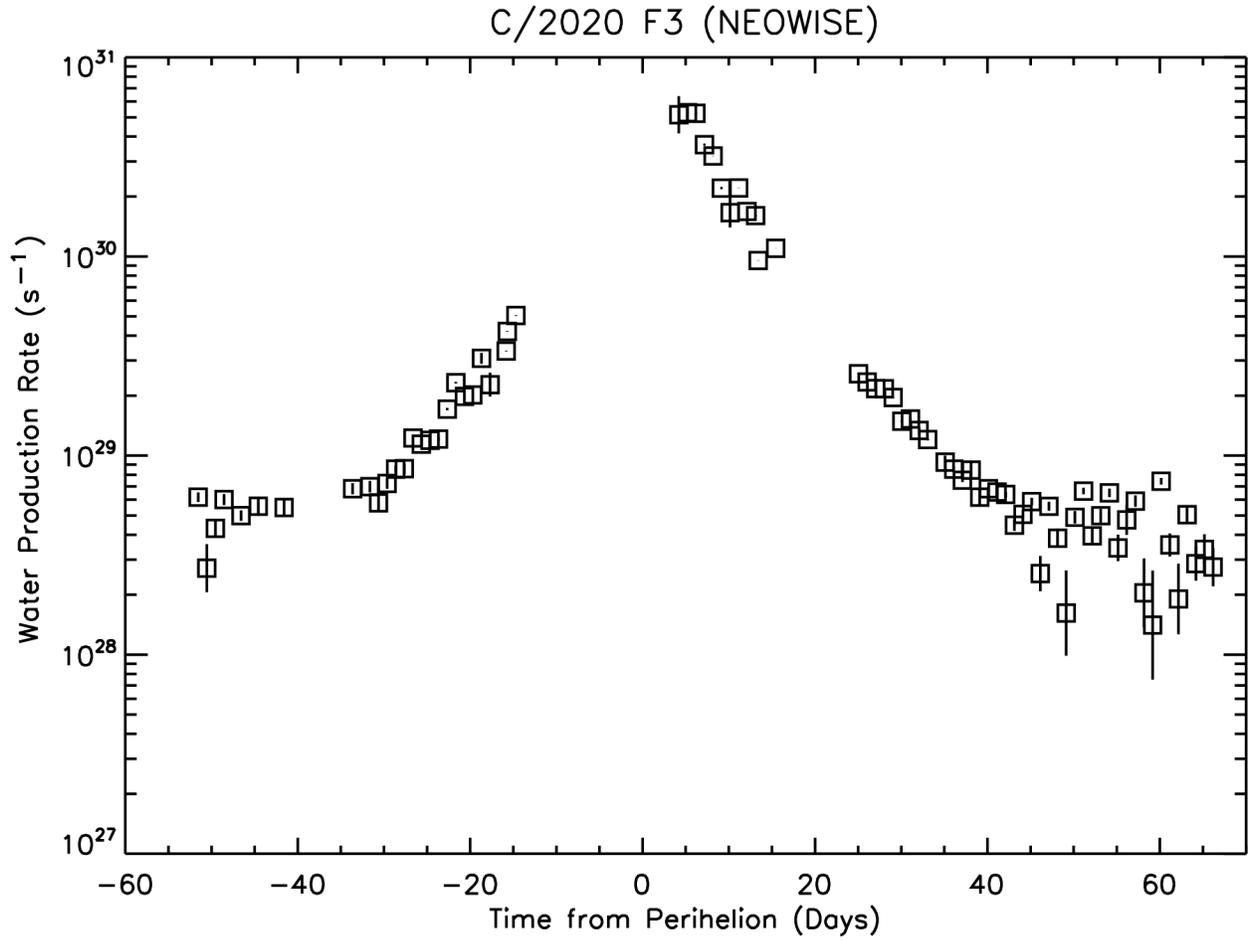

Figure 3. Water production rate in comet C/2020 F3 (NEOWISE) plotted as a function of time in days from perihelion. The error bars indicate the 1-σ uncertainties resulting from noise in the data and the fitting of the coma model and interstellar background.

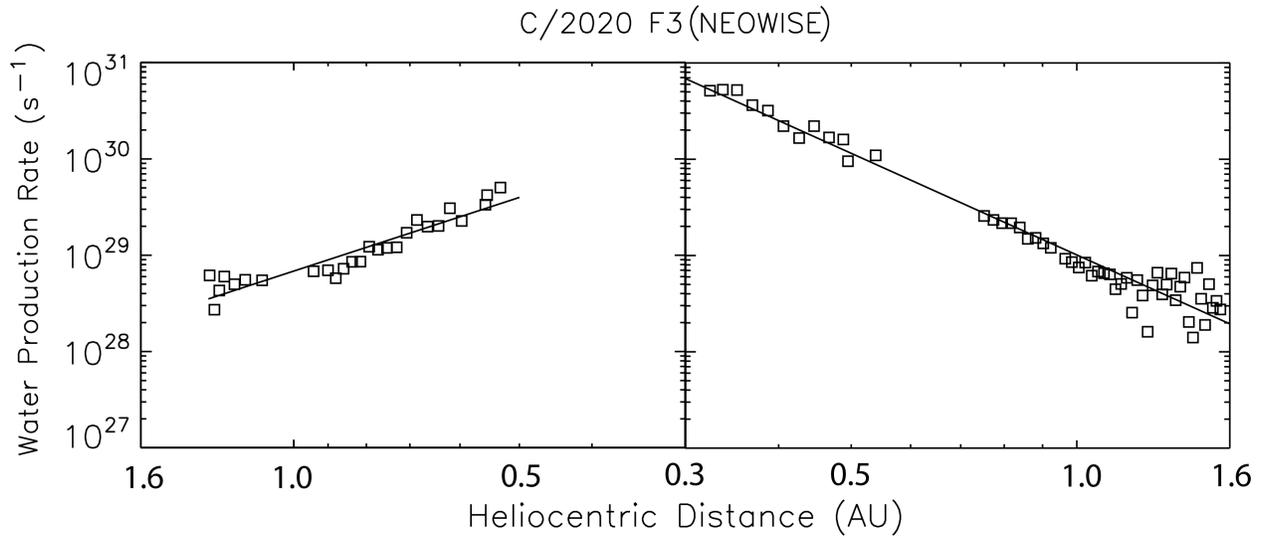

Figure 4. Water production rate of comet C/2020 F3 (NEOWISE) plotted as a function of heliocentric distance. The left half is pre-perihelion and right half is post-perihelion. The straight lines correspond to the power-law fits to the production rate as a function of heliocentric distance for the pre- and post-perihelion legs of the orbit giving values of $(6.9\pm0.5) \times 10^{28}\ r^{-2.5\pm0.2}$ and $(10.1\pm0.5) \times 10^{28}\ r^{-3.5\pm0.1}$, respectively.

Table 1
SOHO/SWAN Observations of C/2020 F3 (NEOWISE) and Water Production Rates

| ΔT (Days) | r (AU) | Δ (AU) | g (s$^{-1}$) | Q (10$^{28}$ s$^{-1}$) | δQ (10$^{28}$ s$^{-1}$) |
|---|---|---|---|---|---|
| −51.553 | 1.295 | 1.603 | 0.001786 | 6.17 | 0.35 |
| −50.552 | 1.276 | 1.604 | 0.001815 | 2.72 | 0.87 |
| −49.552 | 1.257 | 1.604 | 0.001780 | 4.31 | 0.51 |
| −48.552 | 1.238 | 1.604 | 0.001780 | 6.01 | 0.40 |
| −46.575 | 1.200 | 1.605 | 0.001815 | 5.00 | 0.31 |
| −44.580 | 1.161 | 1.604 | 0.001798 | 5.56 | 0.50 |
| −41.580 | 1.102 | 1.603 | 0.001830 | 5.49 | 0.46 |
| −33.634 | 0.941 | 1.590 | 0.001875 | 6.81 | 0.45 |
| −31.634 | 0.900 | 1.584 | 0.001889 | 6.98 | 0.47 |
| −30.634 | 0.879 | 1.581 | 0.001913 | 5.78 | 0.51 |
| −29.637 | 0.858 | 1.577 | 0.001892 | 7.24 | 0.47 |
| −28.636 | 0.836 | 1.573 | 0.001890 | 8.54 | 0.38 |
| −27.637 | 0.815 | 1.569 | 0.001915 | 8.59 | 0.56 |
| −26.636 | 0.793 | 1.564 | 0.001908 | 12.25 | 0.28 |
| −25.663 | 0.772 | 1.560 | 0.001938 | 11.40 | 0.29 |
| −24.662 | 0.751 | 1.554 | 0.001948 | 11.92 | 0.30 |
| −23.663 | 0.729 | 1.548 | 0.001962 | 12.10 | 0.31 |
| −22.664 | 0.707 | 1.542 | 0.001955 | 17.12 | 0.23 |
| −21.664 | 0.685 | 1.535 | 0.001953 | 23.21 | 0.19 |
| −20.664 | 0.663 | 1.528 | 0.001955 | 19.83 | 0.22 |
| −19.692 | 0.641 | 1.520 | 0.001957 | 20.14 | 0.77 |
| −18.692 | 0.619 | 1.511 | 0.001959 | 30.76 | 1.96 |
| −17.692 | 0.597 | 1.502 | 0.001928 | 22.70 | 3.34 |
| −15.819 | 0.555 | 1.483 | 0.001924 | 33.48 | 0.15 |
| −15.693 | 0.552 | 1.482 | 0.001923 | 42.06 | 0.13 |
| −14.693 | 0.530 | 1.471 | 0.001869 | 50.46 | 0.15 |
| 4.203 | 0.324 | 1.011 | 0.001471 | 514.80 | 123.30 |
| 5.200 | 0.337 | 0.976 | 0.001520 | 527.30 | 0.07 |
| 6.174 | 0.352 | 0.944 | 0.001570 | 523.40 | 0.05 |
| 7.174 | 0.369 | 0.911 | 0.001620 | 364.20 | 0.05 |
| 8.172 | 0.387 | 0.880 | 0.001665 | 319.50 | 0.04 |
| 9.145 | 0.406 | 0.852 | 0.001707 | 220.50 | 0.05 |
| 10.145 | 0.426 | 0.824 | 0.001708 | 166.10 | 31.05 |
| 11.144 | 0.446 | 0.799 | 0.001753 | 220.80 | 0.03 |
| 12.116 | 0.467 | 0.777 | 0.001763 | 168.20 | 0.03 |
| 13.116 | 0.488 | 0.756 | 0.001752 | 160.40 | 0.03 |
| 13.412 | 0.495 | 0.751 | 0.001735 | 95.40 | 0.04 |
| 15.440 | 0.539 | 0.719 | 0.001757 | 109.80 | 0.03 |
| 25.043 | 0.752 | 0.727 | 0.001721 | 25.77 | 0.07 |
| 26.043 | 0.774 | 0.741 | 0.001744 | 23.41 | 0.08 |
| 27.043 | 0.795 | 0.757 | 0.001727 | 21.68 | 0.09 |
| 28.044 | 0.817 | 0.775 | 0.001720 | 21.66 | 0.08 |
| 29.064 | 0.839 | 0.795 | 0.001715 | 19.55 | 0.07 |
| 30.064 | 0.860 | 0.816 | 0.001727 | 14.88 | 0.10 |
| 31.064 | 0.881 | 0.839 | 0.001765 | 15.26 | 0.20 |
| 32.073 | 0.902 | 0.863 | 0.001757 | 13.36 | 0.52 |
| 33.073 | 0.923 | 0.888 | 0.001745 | 12.02 | 0.11 |
| 35.092 | 0.965 | 0.940 | 0.001765 | 9.27 | 0.72 |
| 36.092 | 0.985 | 0.968 | 0.001784 | 8.54 | 0.26 |
| 37.092 | 1.006 | 0.996 | 0.001777 | 7.52 | 0.16 |
| 38.102 | 1.026 | 1.025 | 0.001763 | 8.45 | 0.76 |
| 39.102 | 1.047 | 1.054 | 0.001760 | 6.17 | 0.69 |
| 40.102 | 1.067 | 1.083 | 0.001770 | 6.81 | 0.14 |
| 41.121 | 1.087 | 1.114 | 0.001739 | 6.56 | 0.18 |
| 42.121 | 1.107 | 1.144 | 0.001745 | 6.37 | 0.19 |
| 43.121 | 1.126 | 1.175 | 0.001732 | 4.47 | 0.30 |
| 44.121 | 1.146 | 1.206 | 0.001720 | 5.06 | 0.25 |
| 45.131 | 1.166 | 1.237 | 0.001721 | 5.87 | 0.26 |
| 46.131 | 1.185 | 1.268 | 0.001728 | 2.55 | 0.58 |
| 47.131 | 1.204 | 1.300 | 0.001688 | 5.56 | 0.30 |
| 48.131 | 1.224 | 1.331 | 0.001693 | 3.84 | 0.45 |
| 49.131 | 1.243 | 1.363 | 0.001686 | 1.62 | 1.03 |
| 50.147 | 1.262 | 1.395 | 0.001688 | 4.89 | 0.36 |
| 51.147 | 1.281 | 1.427 | 0.001694 | 6.63 | 0.25 |
| 52.147 | 1.300 | 1.459 | 0.001682 | 3.95 | 0.43 |
| 53.147 | 1.318 | 1.491 | 0.001675 | 4.98 | 0.40 |
| 54.148 | 1.337 | 1.523 | 0.001645 | 6.49 | 0.33 |
| 55.147 | 1.355 | 1.554 | 0.001661 | 3.44 | 0.57 |
| 56.157 | 1.374 | 1.587 | 0.001672 | 4.75 | 0.89 |
| 57.157 | 1.392 | 1.619 | 0.001660 | 5.90 | 0.40 |
| 58.157 | 1.411 | 1.650 | 0.001667 | 2.04 | 1.00 |
| 59.157 | 1.429 | 1.683 | 0.001655 | 1.41 | 1.23 |

| ΔT | r | Δ | g | Q | δQ |
|---|---|---|---|---|---|
| 60.157 | 1.447 | 1.714 | 0.001676 | 7.43 | 0.28 |
| 61.157 | 1.465 | 1.746 | 0.001669 | 3.55 | 0.50 |
| 62.157 | 1.483 | 1.778 | 0.001653 | 1.90 | 0.95 |
| 63.157 | 1.501 | 1.809 | 0.001646 | 5.05 | 0.50 |
| 64.174 | 1.519 | 1.841 | 0.001635 | 2.87 | 0.63 |
| 65.157 | 1.536 | 1.872 | 0.001651 | 3.38 | 0.64 |
| 66.174 | 1.554 | 1.903 | 0.001644 | 2.75 | 0.68 |

Notes to Table 1.

ΔT (Days from Perihelion July 3.679, 2020)

r: Heliocentric distance (AU)

Δ: SOHO/Comet distance (AU)

g: Solar Lyman-α g-factor (photons s$^{-1}$) at 1 AU

Q: Water production rates for each image ($10^{28}$ s$^{-1}$)

δQ: internal 1-sigma uncertainties ($10^{28}$ s$^{-1}$)